\documentclass[10pt,preprint]{aastex} 
 
\begin{document}

\title{The Megamaser Cosmology Project. V. An Angular Diameter Distance to NGC 6264 at 140 Mpc}                                                                  
 
\author{C. Y. Kuo\altaffilmark{1}, J. A. Braatz\altaffilmark{2}, M. J. Reid\altaffilmark{3},
K.Y. Lo\altaffilmark{2}, J. J. Condon\altaffilmark{2}, C. M. V. Impellizzeri\altaffilmark{2},
C. Henkel\altaffilmark{4,5}}
 
\affil{\altaffilmark{1}Academia Sinica Institute of Astronomy and Astrophysics, P.O. Box 23-141, Taipei 10617, Taiwan }                                              
\affil{\altaffilmark{2}National Radio Astronomy Observatory, 520
Edgemont Road, Charlottesville, VA 22903, USA}                          
\affil{\altaffilmark{3}Harvard-Smithsonian Center for Astrophysics,
60 Garden Street, Cambridge, MA 02138, USA}                             
\affil{\altaffilmark{4}Max-Planck-Institut f\"ur Radioastronomie,
Auf dem H\"ugel 69, 53121 Bonn, Germany}
\affil{\altaffilmark{5}Astronomy Department, Faculty of Science, King Abdulaziz University, P.O. Box 80203,
Jeddah, Saudi Arabia}

\begin{abstract} 

We present the direct measurement of the Hubble constant, yielding the direct measurement of the angular-diameter distance to NGC 6264 using the H$_{2}$O megamaser technique. Our measurement is based on sensitive observations of the circumnuclear megamaser disk from four observations with the Very Long Baseline Array, the Green Bank Telescope and the Effelsberg Telescope. We also monitored the maser spectral profile for 2.3 years using the Green Bank Telescope to measure accelerations of maser lines by tracking their line-of-sight velocities as they change with time. The measured accelerations suggest that the systemic maser spots have a significantly wider radial distribution than in the archetypal megamaser in NGC 4258. We model the maser emission as arising from a circumnuclear disk with orbits dominated by the central black hole. The best fit of the data gives a Hubble constant of $H_{0} =$ 68$\pm$9 km~s$^{-1}$~Mpc$^{-1}$, which corresponds to an angular-diameter distance of 144$\pm$19 Mpc. In addition, the fit also gives a mass of the central black hole of (3.09$\pm$0.42)$\times$10$^{7}$ $M_{\odot}$. The result demonstrates the feasibility of measuring distances to galaxies located well into the Hubble flow by using circumnuclear megamaser disks. 

\end{abstract} 
 

\section{Introduction}
Cosmology research has entered a new era since the discovery of the acceleration of the Universe (Perlmutter et al. 1999; Riess et al. 1998). ``Dark Energy" (DE), which has negative pressure and accounts for 73\% of the total energy density of the Universe, is currently the best candidate to explain the cosmic acceleration, and understanding its nature has become one of the most important problems in modern astronomy and physics. 

There are several promising methods to explore dark energy with high accuracy via its equation-of-state parameter $w$ (Frieman, Turner \& Huterer 2008). As pointed out by Hu (2005), among all observables for probing DE in light of the Microwave Background Radiation (CMB), $w$ is most sensitive to variations in $H_{0}$. Hu (2005) concluded that the single most important complement to the CMB for measuring $w$ at $z$$\sim$0.5 is a determination of the Hubble constant (at $z$$\sim$0) to better than a few percent. This insight forms the fundamental motivation for the Megamaser Cosmology project (MCP; Reid et al. 2009 (Paper I); Braatz et al. 2010 (Paper II); Kuo et al. 2011 (Paper III); and Reid et al. 2012 (Paper IV)), which aims to determine $H_{0}$ to 3\% accuracy. 

The key to a precise and direct determination of the Hubble constant is to measure accurate distances to galaxies well into the Hubble flow (i.e. $\geq$ 30 Mpc). The galaxies must be distant to reduce the contribution of the uncertainty coming from peculiar velocities relative to a smooth Hubble flow. Among all the approaches to measure accurate distances directly, the megamaser method pioneered by the study of NGC 4258 (Herrnstein et al. 1999) has proven to yield accurate and direct distance measurements to galaxies beyond our Local Group. The methodology of the megamaser technique for distance determination can be found in Paper I, Paper II, and the modified version of the techinique that directly measure the Hubble constant can be found Paper IV. The application of the megamaser technique for measuring black hole mass can be found in Paper III.

In the MCP, we aim to measure the Hubble constant with $\sim$10\% or better accuracy from each of about 10 galaxies, thereby determining $H_{0}$ to $\sim$3\% after averaging the results. The first galaxy measured by the MCP was UGC 3789 (Paper I, Paper II \& Paper IV). A preliminary analysis using a simple model of the maser disk in Paper II gave a Hubble constant of 69$\pm$11 Mpc (16\% accuracy), and the overall uncertainty of the Hubble constant has recently been further reduced to $\sim$10\% when new data and better modeling were included (Paper IV). In this paper, we present the direct Hubble constant measurement from NGC 6264, the first galaxy beyond 100 Mpc measured with the megamaser method. In section 2, we present our VLBI and single-dish observations. In section 3, we show the analysis of the centripetal accelerations of the masers in NGC 6264. The Hubble constant and distance determinations are presented in section 4. In section 5, we discuss the challenges of applying the maser technique to distant galaxies. Finally, we summarize the results in section 6.

\section{Observations and Data Reduction} 
\subsection{VLBI Data}
We observed NGC 6264 with four tracks of VLBI observations between 2009 and 2010 using the Very Long Baseline Array (VLBA), augmented by the 100-m Robert C. Byrd Green Bank Telescope (GBT)\footnote{The VLBA and the GBT are facilities of the National Radio Astronomy Observatory, which is operated by the Associated Universities, Inc. under a cooperative agreement with the National Science Foundation (NSF).} and the 100-m Effelsberg telescope\footnote{The Effelsberg 100-m telescope is a facility of the Max-Planck-Institut f\"ur Radioastronomie}. Two of the four tracks (BB261F and BB261H) were presented in Kuo et al. (2011), and in this paper we present the observations BB261K and BB261Q. Our final VLBI image includes the data from all four tracks. The observing parameters for these four VLBI tracks are shown in Table 1. 
\begin{deluxetable}{lcllc} 
\tabletypesize{\scriptsize} 
\tablewidth{0 pt} 
\tablecaption{Observing Parameters} 
\tablehead{ 
\colhead{Experiment}               & \colhead{}         & 
\colhead{}              & 
\colhead{Synthesized Beam}        & \colhead{Sensitivity}           
       \\ 
\colhead{Code}           & \colhead{Date}        & \colhead{Antennas\tablenotemark{a}}   
&                                                                       
\colhead{(mas x mas,deg)\tablenotemark{b}}  & \colhead{(mJy)\tablenotemark{c}}    
           } 
\startdata 
BB261F   & 2009 Apr 06  & VLBA, GB, EB  & 0.93$\times$0.38,$-$24.8&
0.5  \\                                                   
BB261H   & 2009 Apr 18  & VLBA, GB, EB  & 1.02$\times$0.29,$-$14.7&0.3
 \\        
BB261K   & 2009 Nov 25  & VLBA, GB, EB, VLA-27  & 1.10$\times$0.30,$-$9.7&0.5
 \\ 
BB261Q   & 2010 Jan 15  & VLBA, GB, EB  & 0.80$\times$0.24,$-$11.8&0.5
 \\                                                    
\enddata 
\tablecomments{The observing parameters for NGC 6264. All observations were conducted in the self-calibration mode.}
\tablenotetext{a}{VLBA: Very Long Baseline Array; GB: The Green Bank
Telescope of NRAO; EB: Max-Planck-Institut f\"{u}r Radioastronomie
100 m antenna in Effelsberg, Germany; VLA: The Very Large Array of NRAO. VLA-27 is the phased VLA that involves all 27 antennas in the phased array.}                                  
\tablenotetext{b}{The average FWHM beam size and position angle (PA; measured east of north) at the frequency of systemic masers. }                                   
\tablenotetext{c}{The sensitivities here are the rms noise measured in a channel map with a channel width of 3.6 km~s$^{-1}$.}                                           
\end{deluxetable} 

The observation and data reduction procedures for BB261K and BB261Q are essentially the same as described in section 2.2 of Paper III. Here, we only discuss the slightly different approach to self-calibration used for the two observations presented in this paper. In the previous observations of NGC 6264 (BB261F and BB261H), in order to achieve sufficient sensitivity in the self-calibration process, we required a single maser line (linewidth$\sim$2 km~s$^{-1}$; the channel width is 3.6 km~s$^{-1}$) with flux density above $\sim$80 mJy, or $>$7 maser lines in a single intermediate frequency (IF) band with flux densities above $\sim$30 mJy. In the latter case, we achieved a sufficient signal-to-noise ratio (SNR) by averaging multiple maser lines in narrow ranges of both velocity and spatial position to perform phase calibration. Note that while the multiple maser lines used in the self-calibration are slightly offset on the sky, the position differences among the spots in a single IF band are usually only $\sim$1/10 of the typical synthesized beam of the observation. In this case, we simply averaged the maser signals to enhance the SNR. Based on simulations, the phase shifts caused by such small position differences do not lead to noticeable systematic error in the self-calibration.

In BB261K and BB261Q, the maser lines were generally weaker than those in the previous observations on NGC 6264, and averaging multiple maser lines in a single IF band was not adequate to achieve sufficient SNR for self-calibration. One way to overcome this difficulty is to utilize all maser lines with flux densities above $\sim$20 mJy in \emph{all} IF bands for self-calibration. Since the positional differences of maser spots in different IF bands can be comparable to the synthesized beam, we first removed the phase shifts due to different maser locations by dividing the data by complex visibilities calculated from a model of the maser spot distribution. The model division effectively moves all maser spots in consideration to the phase center of the observation, and since these masers now have approximately the same phase, averaging the maser lines in the calibration process will increase the SNR directly. 

In order to choose the model of maser locations, the best approach is to use maser positions made by calibrating phase with a background quasar (i.e. phase-referencing observations). If available, observations from another epoch when the masers were stronger and self-calibration was successful can also be used. Following this guideline, for BB261K and BB261Q, we used the map from BB261F and BB261H as our model before performing self-calibration. Once the phase shift removal and line averaging are done, we treat the masers as a single maser spot, and follow the same calibration procedure for solving the atmospheric phase, solution editing, and imaging as described in Kuo et al. 2011. Finally, we combine the maser map obtained from BB261K and BB261Q with the one from BB261F and BB261H on the image plane by aligning the maser disks using the average positions of the systemic maser features with velocities between 10191.1 km~s$^{-1}$ and 10198.3 km~s$^{-1}$ as the reference. We present the combined data in Table 2.

\begin{deluxetable}{lrrrrrc} 
\tablewidth{0 pt} 
\tablecaption{Data for NGC 6264} 
\tablehead{ 
\colhead{$V_{\rm op}$\tablenotemark{a}} & \colhead{$\Theta_{x}$\tablenotemark{b}}  
& \colhead{$\sigma_{\Theta_{x}}$\tablenotemark{b}}  & \colhead{$\Theta_{y}$\tablenotemark{b}}   
& \colhead{$\sigma_{\Theta_{y}}$\tablenotemark{b}}  & \colhead{$A$\tablenotemark{c}}  & \colhead{$\sigma_{A}$\tablenotemark{c}} 
\\                                                                      
\colhead{(km~s$^{-1}$)}  &\colhead{(mas)}  
&\colhead{(mas)}         &\colhead{(mas)}         
& \colhead{(mas)}  & \colhead{(km~s$^{-1}$~yr$^{-1}$)} & \colhead{(km~s$^{-1}$~yr$^{-1}$)}        } 
\startdata 
10918.33  &  0.397 &  0.0056 &  $-$0.033 & 0.0141 &  ... & ...\\
10914.71  &  0.403 &  0.0065 &   0.011 & 0.0154 &  ... & ...\\
10911.09  &  0.407 &  0.0048 &  $-$0.005 & 0.0114 &  0.11 & 0.60\\
10907.47  &  0.399 &  0.0069 &   0.026 & 0.0171 &  ... & ...\\
10853.15  &  0.487 &  0.0037 &  $-$0.011 & 0.0091 &  ... & ...\\
10849.53  &  0.494 &  0.0036 &   0.003 & 0.0089 &  0.00 & 0.12\\
10845.91  &  0.501 &  0.0021 &  $-$0.027 & 0.0055 &  0.16 & 0.06\\
10842.29  &  0.504 &  0.0018 &  $-$0.023 & 0.0049 & $-$0.04 & 0.05\\
10838.67  &  0.508 &  0.0042 &  $-$0.024 & 0.0118 & $-$0.24 & 0.07\\
10824.19  &  0.537 &  0.0071 &  $-$0.039 & 0.0159 &  0.27 & 0.34\\
10809.70  &  0.550 &  0.0037 &  $-$0.028 & 0.0092 &  ... & ...\\
10806.08  &  0.561 &  0.0034 &  $-$0.053 & 0.0090 &  ... & ...\\
10802.46  &  0.572 &  0.0017 &  $-$0.031 & 0.0044 & $-$0.09 & 0.08\\
10798.84  &  0.577 &  0.0013 &  $-$0.040 & 0.0034 & $-$0.27 & 0.11\\
10793.40  &  0.582 &  0.0014 &  $-$0.028 & 0.0039 & $-$0.05 & 0.11\\
10787.97  &  0.596 &  0.0046 &  $-$0.032 & 0.0125 &  ... & ...\\
10780.73  &  0.614 &  0.0031 &  $-$0.045 & 0.0075 & $-$0.13 & 0.02\\
10773.49  &  0.632 &  0.0058 &  $-$0.025 & 0.0149 &  ... & ...\\
10766.25  &  0.654 &  0.0051 &  $-$0.046 & 0.0136 & $-$0.26 & 0.03\\
10762.63  &  0.673 &  0.0055 &  $-$0.088 & 0.0142 &  ... & ...\\
10755.38  &  0.664 &  0.0044 &  $-$0.071 & 0.0117 & $-$0.06 & 0.09\\
10751.76  &  0.672 &  0.0020 &  $-$0.052 & 0.0052 & $-$0.07 & 0.04\\
10748.14  &  0.677 &  0.0035 &  $-$0.084 & 0.0099 &  0.02 & 0.14\\
10243.36  &  0.005 & 0.0064  & $-$0.006 & 0.0154 & 4.43  & 0.95\\
10239.75  &  0.013 & 0.0039  &  0.002 & 0.0108  & 4.43  & 0.95\\
10236.15  &  0.012 & 0.0033  &  0.001 & 0.0093  & 4.43  & 0.95\\
10232.54 &   0.008 & 0.0062  &  0.015 & 0.0147  & 4.43  & 0.95\\
10218.11 &   0.007 & 0.0032  & $-$0.008 & 0.0080  & 1.55 & 0.46\\
10203.69 &  $-$0.008 & 0.0028  &  0.007 & 0.0075  & 0.75 & 0.14\\
10200.08 &  $-$0.010 & 0.0026  & -0.001 & 0.0080  & 1.07 & 0.32\\
10196.47 &  $-$0.009 & 0.0014  &  0.004 & 0.0039  & 1.07 & 0.32\\
10192.87 &  $-$0.014 & 0.0012  &  0.004 & 0.0035  & 1.07 & 0.32\\
10189.26 &  $-$0.015 & 0.0017  &  0.007 & 0.0047  & 1.07 & 0.32\\
10185.65 &  $-$0.019 & 0.0028  &  0.013 & 0.0073  & 1.07 & 0.32\\
 9787.19 &  $-$1.105 & 0.0054  &  0.321 & 0.0159  & 0.02 & 0.03\\
 9682.93 &  $-$0.702 & 0.0034  &  0.083 & 0.0087  & 0.00 & ...\\
 9679.34 &  $-$0.678 & 0.0015  &  0.073 & 0.0040  & 0.29 & 0.08\\
 9675.75 &  $-$0.675 & 0.0025  &  0.087 & 0.0065  & ...  &...\\
 9672.15 &  $-$0.689 & 0.0060  &  0.088 & 0.0139  & ...  &...\\
 9661.37 &  $-$0.645 & 0.0054  &  0.047 & 0.0147  & ...  &...\\
 9657.77 &  $-$0.638 & 0.0058  &  0.070 & 0.0139  & ...  &...\\
 9654.18 &  $-$0.618 & 0.0064  &  0.064 & 0.0159  & ...  &...\\
 9650.58 &  $-$0.615 & 0.0043  &  0.081 & 0.0116  & ...  &...\\
 9646.99 &  $-$0.611 & 0.0051  &  0.063 & 0.0125  & ...  &...\\
 9643.39 &  $-$0.591 & 0.0037  &  0.056 & 0.0100  & ...  &...\\
 9639.80 &  $-$0.582 & 0.0062  &  0.018 & 0.0152  & ...  &...\\
 9636.20 &  $-$0.567 & 0.0055  &  0.029 & 0.0129  &$-$0.04 & 0.07\\
 9632.61 &  $-$0.571 & 0.0055  &  0.071 & 0.0144  & 0.22 & 0.08\\
 9629.01 &  $-$0.566 & 0.0052  &  0.062 & 0.0142  & 0.00 & 0.20\\
 9614.63 &  $-$0.505 & 0.0046  &  0.043 & 0.0108  &$-$0.12 & 0.06\\
 9611.04 &  $-$0.499 & 0.0044  &  0.036 & 0.0105  & 0.11 & 0.10\\
 9649.42 &  $-$0.610 & 0.0052  &  0.071 & 0.0151  & ...  &...\\
 9645.83 &  $-$0.598 & 0.0054  &  0.070 & 0.0155  & ...  &...\\
 9642.24 &  $-$0.595 & 0.0040  &  0.055 & 0.0113  & ...  &...\\
 9635.06 &  $-$0.565 & 0.0064  &  0.044 & 0.0158  &$-$0.04 & 0.07\\
 9631.47 &  $-$0.571 & 0.0051  &  0.091 & 0.0157  & 0.22 & 0.08\\
 9613.52 &  $-$0.502 & 0.0044  &  0.034 & 0.0112  &$-$0.06 & 0.07\\
 9609.92 &  $-$0.514 & 0.0043  &  0.042 & 0.0123  & ...  &...\\
 9584.79 &  $-$0.491 & 0.0044  &  0.047 & 0.0133  & ...  &...\\
 9581.20 &  $-$0.478 & 0.0043  &  0.032 & 0.0132  & ...  &...\\
 9577.61 &  $-$0.476 & 0.0048  &  0.047 & 0.0133  & ...  &...\\
 9574.02 &  $-$0.476 & 0.0054  &  0.042 & 0.0155  & ...  &...\\
 9566.83 &  $-$0.460 & 0.0045  &  0.007 & 0.0129  & ...  &...\\
 9563.24 &  $-$0.451 & 0.0038  &  0.010 & 0.0110  & ...  ...\\
 9538.11 &  $-$0.397 & 0.0051  &  0.001 & 0.0145  & 0.15  &0.20\\
 9534.52 &  $-$0.408 & 0.0033  &  0.020 & 0.0091  &$-$0.44  &0.17\\
 
\enddata 
\tablecomments{ }
\tablenotetext{a}{Velocity referenced to the LSR and using the optical definition (no relativistic corrections).}  
\tablenotetext{b}{East-west and north-south position offsets and
uncertainties measured relative to the average position of the
systemic masers from the BB261K and BB261Q data. Position uncertainties reflect fitted random errors
only. }
\tablenotetext{c}{Measured or estimated acceleration and its uncertainty for each maser component. }    
\end{deluxetable}

\subsection{GBT monitoring}
We observed the H$_{2}$O masers in NGC 6264 with the GBT at 20 epochs between 2008 November 21 and 2011 March 2. Except during the summer months when the humidity makes observations at 22 GHz inefficient, we took a spectrum on a monthly timescale. For these observations, we follow the same observing settings and data reduction procedures as in Braatz et al. (2010). Table 3 shows the observing date and sensitivity for each observation. A representative H$_{2}$O maser spectrum for NGC 6264 can be found in Figure 1 in Kuo et al. 2011.

\begin{deluxetable}{clcccc} 
\tabletypesize{\scriptsize} 
\tablewidth{0 pt} 
\tablecaption{GBT Observing dates and sensitivities for NGC 6264} 
\tablehead{ 
\colhead{Epoch} & \colhead{Date}      & \colhead{Day Number}
 & \colhead{T$_{sys}$ (K)}  &                                                 
\colhead{rms Noise (mJy)} & Period}     
\startdata 
0 & 2008 November 21  & 0  & 30.9 &  1.7 & A \\
1 & 2009 January 16  & 56  & 28.2 & 1.3  & A\\
2 & 2009 February 3  & 74  & 35.9 &1.3  & A \\
3 & 2009 March 4     & 103  & 31.9  &1.2 & A\\
4 & 2009 March 31    & 130  & 37.8 &1.4 & A\\
5 & 2009 May 13      & 172  & 41.5 &1.7 & A\\
6 & 2009 November 7  & 351  & 44.3 &1.7 & B\\
7 & 2009 December 12  & 386  & 27.4  &1.5 & B \\
8 & 2010 January 11  & 416  & 26.3 &1.1 & B\\
9 & 2010 February 9  & 445  & 39.0 &1.5 & B\\
10 & 2010 March 7  & 471  & 28.3 &1.5 & B\\
11 & 2010 April 13  & 507  &36.8  &2.2 & B \\
12 & 2010 May 10  & 534  & 28.2 &1.2 & B\\
13 & 2010 July 2  & 588  & 40.3 &2.2 & B\\
14 & 2010 October 30  & 708  & 40.2 &1.7 & C\\
15 & 2010 November 26  & 735  & 38.4 &1.5 & C\\
16 & 2010 December 24  & 763  & 30.3 &1.5 & C\\
17 & 2011 January 23  & 793  & 31.4 &1.2 & C\\
18 & 2011 February 7  & 808  & 42.7 &1.4 & C\\
19 & 2011 March 2  & 831  & 39.0 &2.3 & C\\
\enddata 
\tablecomments{The rms noise values are calculated without performing Hanning smoothing to the spectra and are based on 0.33 km~s$^{-1}$ channels. We label Period A, B, and C to those times when we have continuous observations on a monthly timescale. These periods are separated by summer months during which the humidity makes observations inefficient. }
      
\end{deluxetable} 

\clearpage  
 
\section{Acceleration Analysis} 

\subsection{Methodology}
Water masers in circumnuclear disks experience centripetal acceleration as they orbit the central black hole. We measure these accelerations by tracking the line-of-sight velocity of each maser line in the spectra obtained by the GBT as they change with time. These acceleration measurements are used along with the positions and velocities of maser spots obtained from the VLBI maps to construct the maser disk model for distance determination described in section 4.

In this paper, we adopted two different methods to determine the accelerations of maser spots, one method for high velocity masers\footnote{The high velocity masers are located near the mid-line of the maser disk, and have radial velocities redshifted or blueshifted from the systemic velocity of the galaxy by $\sim$600 km~s$^{-1}$.} and the other for systemic masers\footnote{The systemic masers refer to the maser components having radial velocities close to the systemic velocity of the galaxy, typically within 100 km~s$^{-1}$.}. We call the first method the \emph{eye-tracking} method. In this method, we first measure the peak velocities of maser spectra at different epochs, and plot the spectral peak velocities as a function of time (e.g. in Fig. 1). We then identify a group of data points within a particular time segment by eye that we believe to be from a single maser line, and fit the trend of the data with straight lines. The slopes of the fitted lines directly give the accelerations. We apply this method to high velocity masers.

The second method is based on a global least-square fitting program (Braatz et al. 2010; Humphreys et al. 2008) that we call \emph{GLOFIT}. Here we fit multiple maser lines from multiple epochs simultaneously, and determine the accelerations of the fitted maser lines in a single step. This is the same method we used for measuring accelerations for UGC 3789 (Paper II) and described in detail as ``Method-1" in Paper IV. We apply this method to systemic masers.   

Among these two methods, we choose the GLOFIT technique for the systemic masers because the GLOFIT method can better handle the line-blending effect of the systemic masers, and determine their accelerations more accurately. On the other hand, since many high velocity maser lines are relatively isolated, and very precise acceleration measurements of these masers are not critical for accurate distance determination, we apply the quicker eye-tracking method to provide estimates of accelerations for the high velocity masers.  

When using the GLOFIT program to measure the accelerations of the systemic masers, it was necessary to use a different strategy for NGC 6264 than for UGC 3789, because the effects of strong blending and low SNR made finding stable and converging solutions more difficult. We mitigated these effects by grouping the masers lines into groups of lines (clumps), in each of which all the maser lines were fitted with a single acceleration parameter. In principle, this approach should help the program to reach convergence more efficiently and stably, because fewer parameters are needed when we enforce a common acceleration. We have found that this is indeed the case by testing this approach with synthetic spectra generated from simulations.

Finally, the applicability of the GLOFIT program, as described above, depends critically on the assumption that masers within the selected velocity range drift at the same rate. Therefore, to avoid systematic errors it is important to examine whether the chosen sections of spectra really satisfy this assumption. The constant acceleration assumption can be checked in several ways, including whether the eye-tracking method gives nearly the same accelerations for maser lines within the chosen spectral region, the persistence of the maser spectral pattern with time, and the reduced $\chi^{2}$ of the fit. We refer interested readers to Kuo 2012 (Ph.D. thesis) for details.

\subsection{Acceleration Fitting for High Velocity Masers}
In Figure 1, we plot the radial velocities of NGC 6264 maser peaks measured by eye as a function of time. For the blueshifted and redshifted masers, we first identify lines that are persistent in time and then fit a straight line to the data to measure the acceleration directly with the eye-tracking method. We estimate the uncertainty of the measurements by scaling the fitting error by the square root of reduced $\chi^{2}$. In Table 4, we list the measured accelerations and uncertainties.

\begin{figure}[ht] 
\begin{center} 
\vspace*{0.0 cm} 
\hspace*{-1.0 cm} 
\includegraphics[angle=0, scale=0.7]{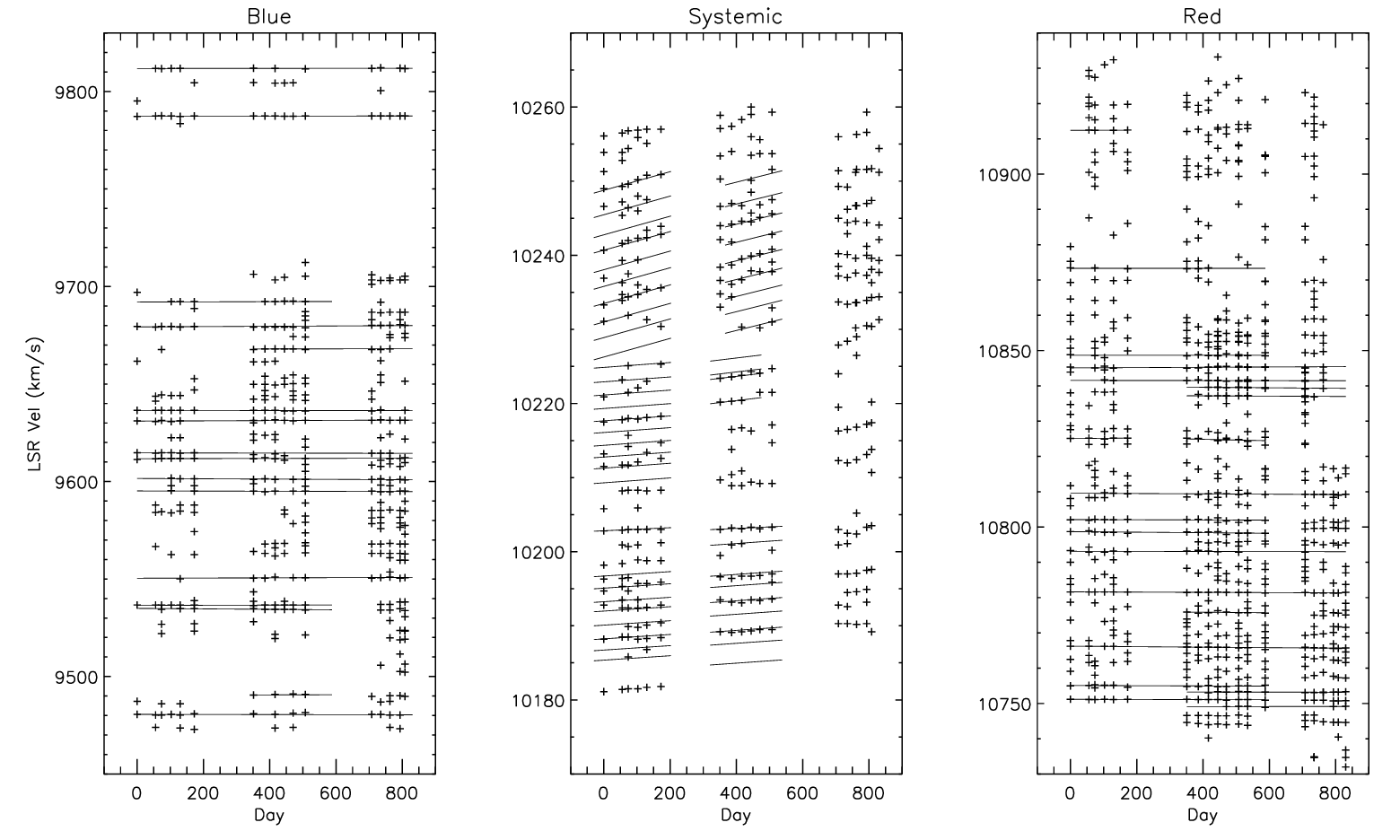}
\vspace*{0.0 cm} 
\caption{Radial velocities of NGC 6264 maser peaks as a function of time (the crosses) along with the fitting results from the eye-tracking method (for the high velocity masers) or from the modified GLOFIT program (for the systemic masers).The data between Day 0 and 200 come from spectra taken in Period A; the data between Day 300 to 600 from spectra in Period B; and the data between Day 700 to 900 are from spectra in Period C. Note that for the acceleration fitting shown in the middle panel, there are some fits in Period B without data. This is because the spectral region for these fits are more significantly blended and it is therefore difficult to identify maser peaks by eye. 
}
\end{center} 
\end{figure}

\begin{deluxetable}{rrc} 
\tabletypesize{\scriptsize} 
\tablewidth{0 pt} 
\tablecaption{Acceleration Measurements for the High Velocity Masers in NGC 6264} 
\tablehead{ 
\colhead{Velocity}      & \colhead{Acceleration}
 & \colhead{$\sigma_{accel}$} \\
\colhead{(km~s$^{-1}$)}      & \colhead{(km~s$^{-1}$~yr$^{-1}$)}
 & \colhead{(km~s$^{-1}$~yr$^{-1}$)}  
}     
\startdata 
9480.57    &  $-$0.09   &   0.07  \\
9490.42    &   0.14   &   0.55  \\
9536.46    &   0.15   &   0.420  \\
9534.93    &  $-$0.44   &   0.17  \\
9550.39    &   0.19   &   0.12  \\
9595.16    &  $-$0.13   &   0.08  \\
9601.54    &  $-$0.28   &   0.08  \\
9611.70    &   0.11   &   0.10  \\
9614.70    &  $-$0.12   &   0.06  \\
9630.99    &   0.23   &   0.08  \\
9636.43    &  $-$0.04   &   0.07  \\
9667.88    &   0.09   &   0.10  \\
9679.29    &   0.29   &   0.08  \\
9692.04    &   0.17   &   0.10  \\
9787.36    &   0.02   &   0.03  \\
9811.77    &   0.05   &   0.06  \\ 
           &          &         \\
10749.05   &   0.02   &   0.14  \\
10751.24   &  $-$0.07   &   0.04  \\ 
10753.28   &  $-$0.03   &   0.08  \\
10755.20   &  $-$0.10   &   0.10  \\
10766.50   &  $-$0.26   &   0.03  \\
10776.50   &  $-$0.30   &   0.42  \\
10781.77   &  $-$0.13   &   0.02  \\
10793.18   &  $-$0.05   &   0.10  \\
10798.96   &  $-$0.27   &   0.11  \\
10802.19   &  $-$0.09   &   0.08  \\
10809.58   &  $-$0.17   &   0.06  \\
10824.78   &   0.27   &   0.34  \\
10826.46   &  $-$0.75   &   0.73  \\
10837.53   &  $-$0.16   &   0.08  \\
10840.21   &  $-$0.27   &   0.06  \\
10841.58   &  $-$0.04   &   0.05  \\
10845.09   &   0.16   &   0.06  \\
10848.68   &   0.00   &   0.11  \\
10873.30   &   0.00   &   0.16  \\
10912.41   &   0.11   &   0.60  \\

\enddata 
\tablecomments{The acceleration measurements for the high velocity masers. The components having velocities higher than 10745 km~s$^{-1}$ are redshifted masers whereas those having velocities lower than 9815 km~s$^{-1}$ are blueshifted masers}
      
\end{deluxetable}

The weighted average accelerations of the redshifted and blueshifted masers are $-$0.06 and 0.01 
km~s$^{-1}$~yr$^{-1}$ respectively. The rms scatter of acceleration of the redshifted and blueshifted masers are both 0.11 km~s$^{-1}$~yr$^{-1}$. The small acceleration and rms scatter indicate that the high velocity masers are close to the mid-line of the accretion disk as expected. 

When assigning the measured accelerations to the corresponding channels in the VLBI datasets, we pay attention to whether the periods over which the accelerations are measured actually bracket the epochs of our VLBI observations. We have two tracks (BB261F \& BB261H) of VLBI observations in Period A (2008 November 21 $-$ 2009 May 13; see Table 3), two tracks (BB261K \& BB261Q) in Period B (2009 November 7 $-$ 2010 July 2) and no VLBI data in Period C (2010 October 30 $-$ 2011 May 2). For those masers persisting over Periods A through C, we assign measured accelerations to the corresponding VLBI channels from images made by averaging all four tracks. For those masers that persist only during Period A or B, we assign the accelerations only to the VLBI channels from observations in their respective periods. For high velocity maser spots without acceleration measurements, we ignore their accelerations in the disk modeling and use only the position-velocity data of these maser spots to constrain the model of NGC 6264.

\subsection{Acceleration Fitting for Systemic Masers}
Figure 2 shows the maser spectra from Epochs 0 through 5 (Period A) and from Epochs 6 through 11 (Period B; see Table 3). We do not show the spectra from Period C because we do not have VLBI data for Period C, and the acceleration measurements for the systemic masers in this period are not used for the distance determination. Since both Periods A and B bracket two tracks of VLBI observations and the accelerations of systemic masers may change with time, we measure acceleration in Periods A and B separately. In cases where the accelerations measured in both periods within a certain velocity range are the same and the corresponding VLBI positions are consistent, we average the accelerations and VLBI positions. For those maser lines that only persist during one period, we only use the VLBI positions at this particular period for these masers in the modeling.    

To measure the accelerations, we break the spectra into distinct velocity ranges, as indicated in Figure 2, and analyze each range individually. 
In particular, we group the systemic masers in Period A into four clumps and those in Period B into five clumps. Note that while this kind of complicated clump grouping was not performed for UGC 3789 in Paper II \& IV, such grouping is important for measuring reliable accelerations for NGC 6264. As mentioned in section 3.1, because of strong blending and low SNR, we need to fit the maser lines in the velocity ranges of interest by enforcing a common acceleration in order to obtain stable and converging solutions. Thus, before fitting accelerations with the GLOFIT program, we must first define spectral regions where the constant acceleration assumption could be satisfied. Based on the acceleration estimates from the eye-tracking method and the degree of persistency of the maser spectral pattern with time, we identify these nine systemic maser clumps (see Figure 2) in which it is likely that the maser lines have approximately the same acceleration. We then analyze each clump individually and show the analysis in the following paragraphs. We present the measured accelerations in Table 5 and plot acceleration measurements as a function of LSR velocity in Figure 3.

\begin{figure}[ht] 
\begin{center} 
\vspace*{0.0 cm} 
\hspace*{0.0 cm} 
\includegraphics[angle=0, scale=0.50]{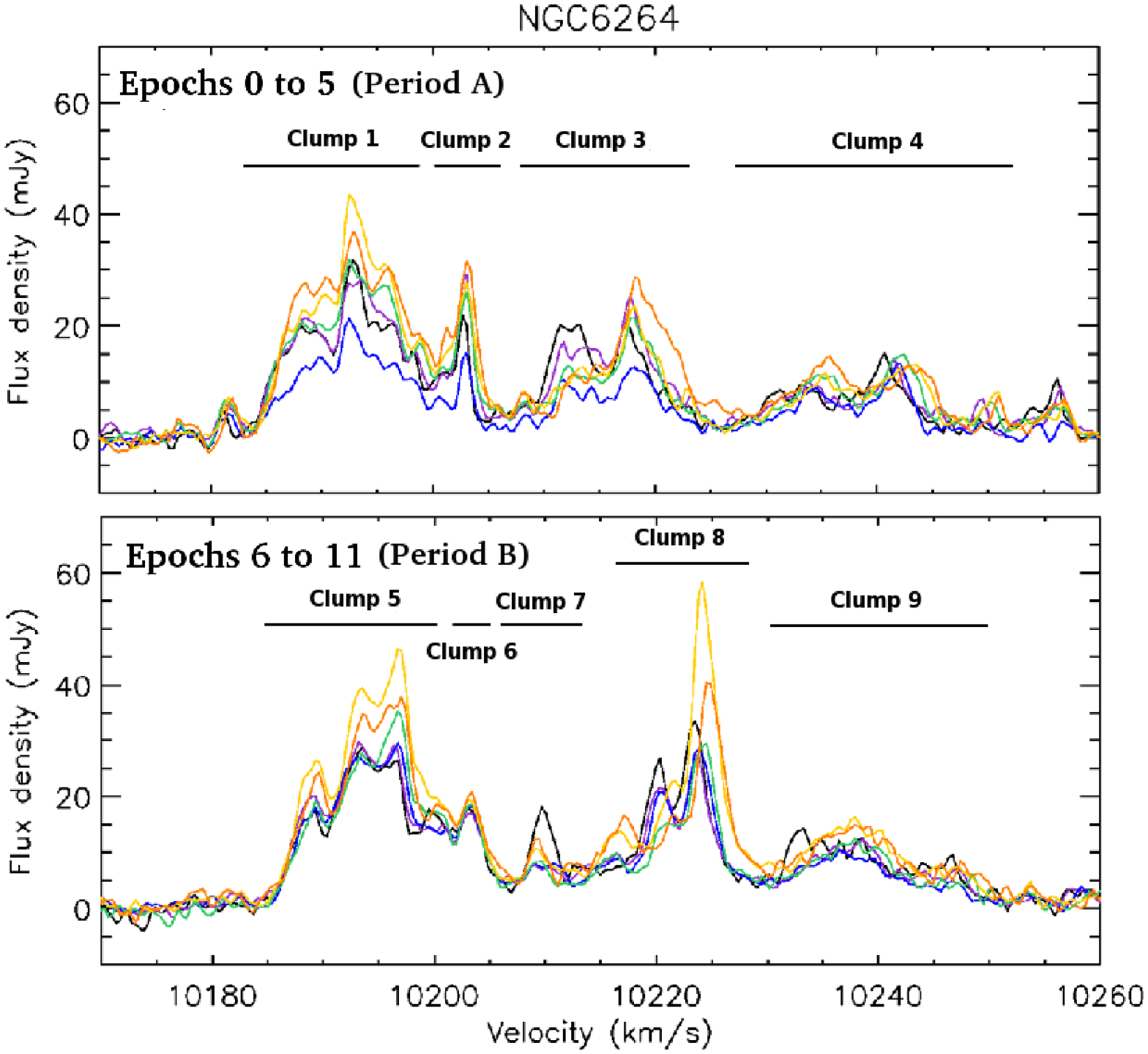}
\vspace*{0.0 cm} 
\caption{The upper panel shows the spectra from epoch 0 through 5 (purple, blue, green, yellow, and orange), and the bottom panel shows the spectra from epoch 6 through 11 (purple, blue, green, yellow, and orange). The velocity ranges of the systemic masers in Period A and B are divided into 7 clumps for the convenience of the acceleration measurement.}
\end{center} 
\end{figure}  
      
\begin{figure}[ht] 
\begin{center} 
\vspace*{0.0 cm} 
\hspace*{0.0 cm} 
\includegraphics[angle=0, scale=0.60]{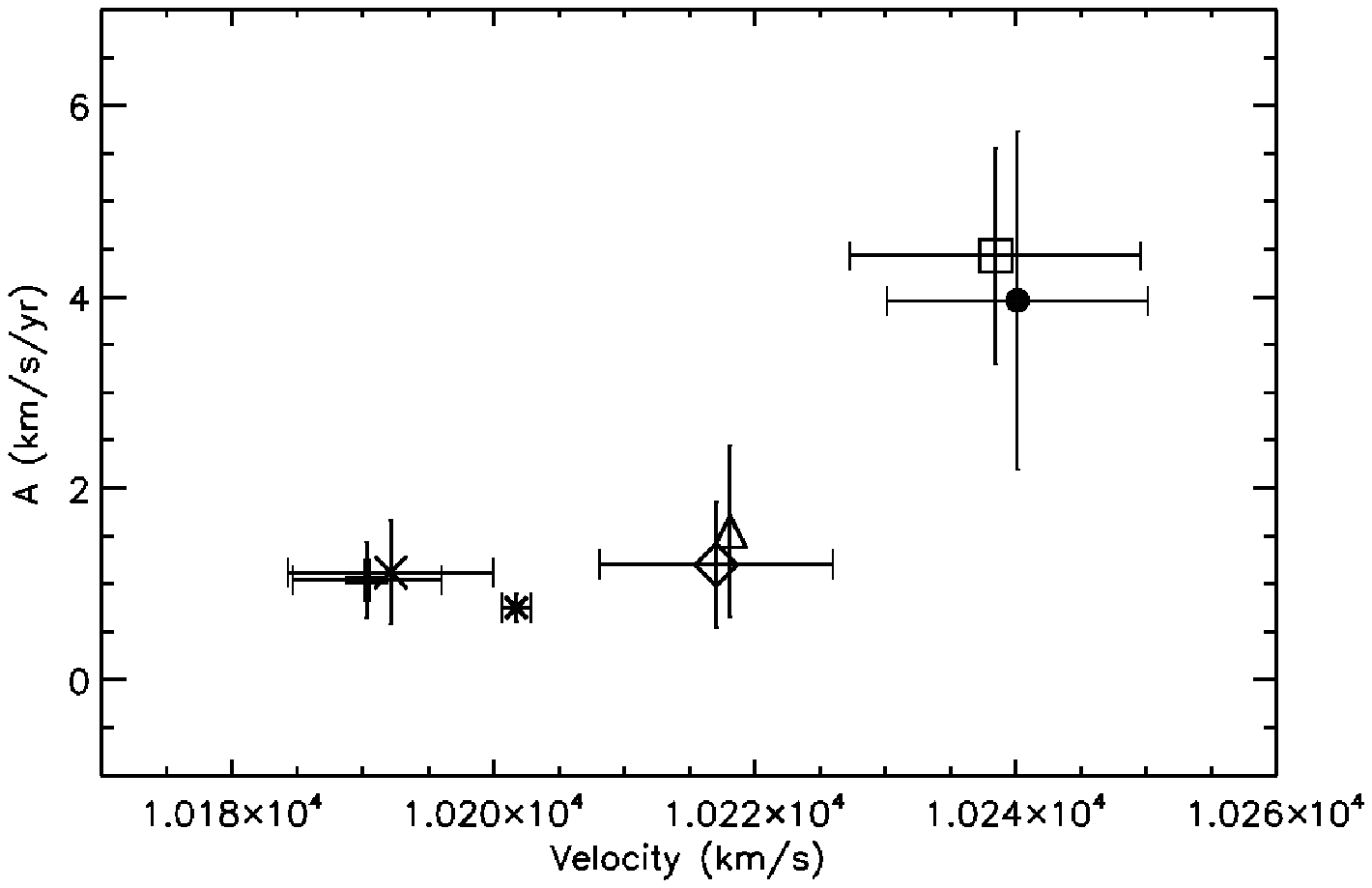}
\vspace*{0.0 cm} 
\caption{Acceleration measurement as a function of LSR velocity for systemic maser features. The different symbols correspond to maser features in Clump 1 (plus sign), Clump 2 (asterisk), Clump 3 (diamond), Clump 4 (square), Clump 5 (cross), and Clump 9 (filled circle), respectively. The triangle shows the dominant maser feature in Clump 3 ($V=$10218.11 km~s$^{-1}$) for which we have a reliable acceleration measurement ($a$$=$1.55$\pm$0.90 km~s$^{-1}$~yr$^{-1}$) from the eye-tracking method. The horizontal bar for each clump shows the velocity range within which we fit the spectra with the constant acceleration assumption. The vertical bars show the measurement errors for the accelerations. Acceleration measurements for masers in Clump 6, 7, and 8 are not plotted because we do not find reliable solutions for these masers.   }
\end{center} 
\end{figure} 

\subsubsection{Clump 1}
The maser lines of Clump 1 reside in a velocity range between 10183.0 to 10197.7 km~s$^{-1}$. Within this velocity range, we fit 8 Gaussian components to the spectra and obtain an acceleration of 1.04$\pm$ 0.40 km~s$^{-1}$~yr$^{-1}$. The reduced $\chi^{2}$ for this fit is 1.144 (153 degrees of freedom).  We over-plot the acceleration measurements from this and the following subsections on the time-velocity plot for the systemic masers (the middle plot of Figure 1); we show an example of the result of the Gaussian decomposition in the acceleration fitting with the masers in Clump 1 in Figure 4.    

\subsubsection{Clump 2}
For the maser lines between 10200.2 to 10203.5 km~s$^{-1}$, we fit two Gaussian components and obtain an acceleration of 0.75$\pm$0.14 km~s$^{-1}$~yr$^{-1}$. The reduced $\chi^{2}$ for the fit is 0.966 (40 degrees of freedom).

\subsubsection{Clump 3}
The maser lines in Clump 3 are apparently drifting and the general pattern of the clump seems to persist with time. However, we couldn't find a good model to fit the data. The best reduced $\chi^{2}$ we can achieve is 1.44, and the fitted acceleration is 1.2$\pm$0.66 km~s$^{-1}$~yr$^{-1}$. The situation does not improve when we only fit subsets of this clump. Based on our experience with the UGC 3789 data, the large reduced $\chi^{2}$ for this kind of situation suggests that some maser features emerge and decay in a time scale shorter than the fitting period, and the fitting program couldn't handle this discontinuity in the velocity drift.  Therefore, to avoid significant systematic uncertainties, for masers in this clump we only use the VLBI channel at $V=$10218.11 km~s$^{-1}$ in the distance determination because it is the only line for which we have a reliable acceleration measurement ($a$$=$1.55$\pm$0.90 km~s$^{-1}$~yr$^{-1}$) from a eye-tracking fit for this clump, and the value is consistent with the one obtained from the GLOFIT method. 

\subsubsection{Clump 4}
Maser lines in Clump 4 have the largest acceleration among the systemic masers in NGC 6264. The drifting of the whole clump can be seen clearly in Figure 2. We fit ten Gaussian components with an average linewidth of 2.4 km~s$^{-1}$ to the masers between 10226.0 to 10252.3 km~s$^{-1}$. The measured acceleration is 4.43$\pm$1.13 km~s$^{-1}$~yr$^{-1}$ with a reduced $\chi^{2}$ of 1.006 (314 degrees of freedom). 

\subsubsection{Clump 5}
The maser lines in Clump 5 reside within nearly the same velocity range as Clump 1, but cover time period B, and the situation is similar. We fit 9 Gaussian components to the spectral region between 10184 to 10201.6 km~s$^{-1}$ and obtain an acceleration of 1.12$\pm$0.54 km~s$^{-1}$~yr$^{-1}$ with a reduced $\chi^{2}$ of 1.009. 

\subsubsection{Clump 6}
The maser lines in Clump 6 reside in the velocity range between 10201.6 to 10205.5 km~s$^{-1}$. We could not find a reliable fit for this maser clump because of severe blending. By comparing the maser spectra from Period A and Period B, we argue that since the line structure remains similar over Periods A and B at this velocity range, we are seeing the same maser lines that appear at slightly different velocities because of the acceleration. This idea is supported by the fact that the VLBI positions of the masers in this velocity range from Periods A and B are the same within the errors. For these reasons, we assume these masers have the same acceleration ($a~=$ 0.75$\pm$0.14 km~s$^{-1}$~yr$^{-1}$) as the masers in Clump 1 at the similar velocity range.

\begin{figure}[ht] 
\begin{center} 
\vspace*{0.0 cm} 
\hspace*{0.0 cm} 
\includegraphics[angle=0, scale=0.8]{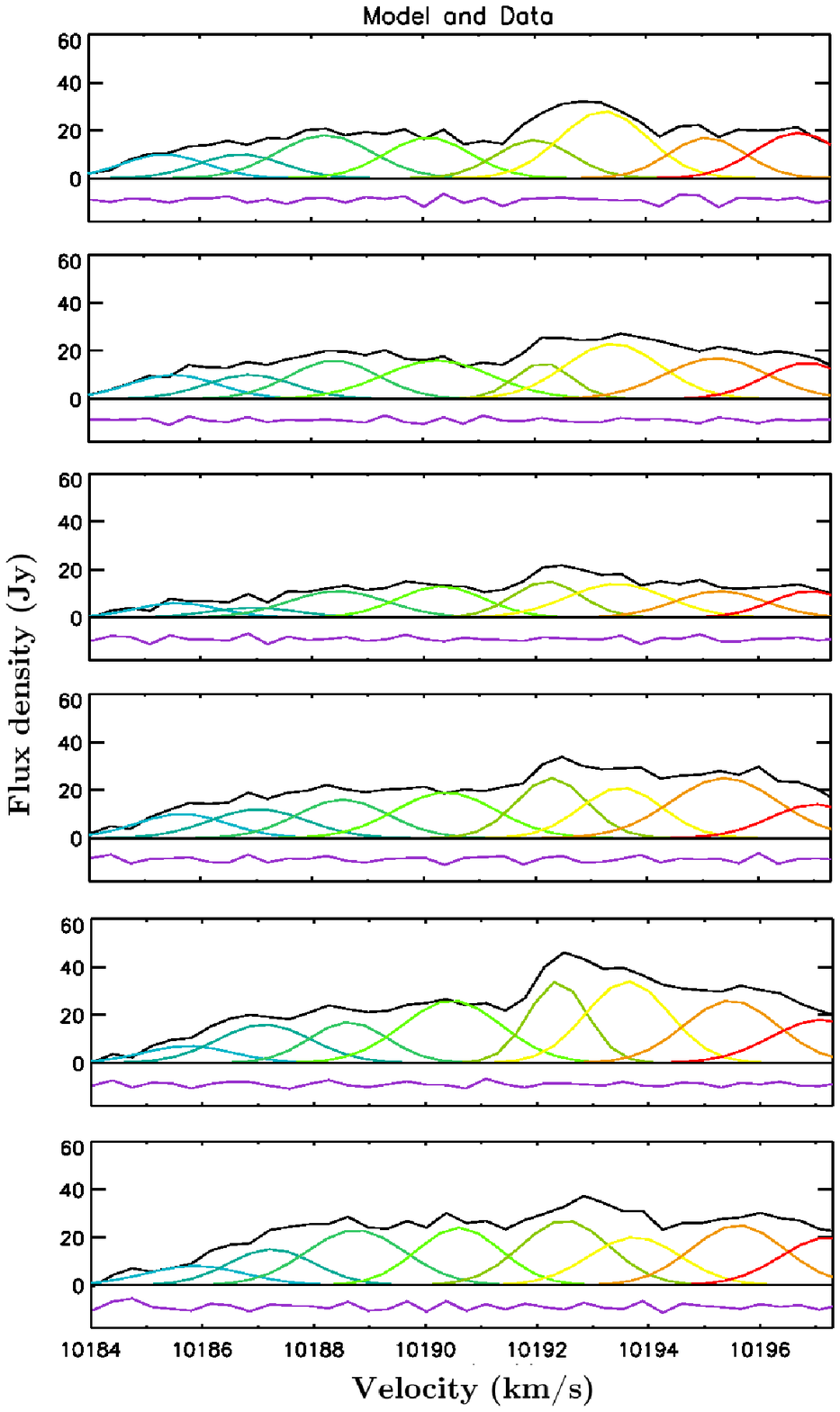}
\vspace*{0.0 cm} 
\caption{An example of the Gaussian decomposition for the acceleration measurement. Here, we fit the masers between 10183.0 and 10197.7 km~s$^{-1}$ (Clump 1; see section 3.3.1). The panels from top to bottom show the spectra (black lines) from epoch 0 through 5. Each of the eight Gaussian components fitted to the data are represented by different colors. The purple curves at the bottom of each panel are the residuals from the fit. Note that the residuals are offset from zero for clarity.}
\end{center} 
\end{figure}

\subsubsection{Clump 7} 
Clump 7 covers a velocity range between 10207 to 10212 km~s$^{-1}$ and the lines within this velocity range
do not show a clear sign of drifting. In addition, the line structure changes substantially over the course of time and we found it difficult to measure a reliable acceleration for this clump. We can measure significantly different accelerations, ranging from 0.4 to 4.2 km~s$^{-1}$~yr$^{-1}$ by using different subsets of the spectra in the fit. This implies that we may see different lines at different epochs. Along with the fact that the VLBI positions for masers in this clump have relatively large error bars, we decided not to include these masers in the modeling.  

\subsubsection{Clump 8}
The acceleration measurement for masers in clump 8 is also difficult as for Clump 7. We couldn't achieve a converging fit. The line structure changes substantially at epoch 10. By inspecting the spectra in Period B along with those in Period C, we discovered that the line at V $\sim$ 10224.1 km~s$^{-1}$ starts to drift at a much higher rate ($a~>$ 7 km~s$^{-1}$~yr$^{-1}$) after epoch 9. This may be caused by newly arising lines from smaller radii in the disk. Because of this complexity, which could introduce significant systematic error in our fitting caused by maser lines with different accelerations, we do not include this group of masers in the Hubble constant determination.  

\begin{deluxetable}{ccccccccc} 
\tabletypesize{\scriptsize} 
\tablewidth{0 pt} 
\tablecaption{Acceleration Measurements for the Systemic Masers in NGC 6264} 
\tablehead{ 
\colhead{Clump}      & \colhead{Velocity Range} & \colhead{Epochs}           & \colhead{Num. of}  & 
\colhead{Linewidth}  & \colhead{Acceleration}   & \colhead{$\sigma_{accel}$} & \colhead{$\chi^{2}_{\nu}$}  &  
\colhead{d.o.f.} \\
\colhead{} & \colhead{(km~s$^{-1}$)} & \colhead{}  & \colhead{Components} & \colhead{(km~s$^{-1}$)}    & \colhead{(km~s$^{-1}$~yr$^{-1}$)}
 & \colhead{(km~s$^{-1}$~yr$^{-1}$)} &  \colhead{}  &  \colhead{}  
}     
\startdata 
1   &  10183.0 $-$ 10197.7  &  0 $-$ 5  & 8  &  1.9  &  1.04  & 0.40  &  1.144  &  153  \\
2   &  10200.2 $-$ 10203.5  &  0 $-$ 5  & 2  &  1.9  &  0.75  & 0.14  &  0.966  &  40  \\
3   &  10209.1 $-$ 10228.0  &  1 $-$ 5  & 11 &  1.7  &  1.20  & 0.66  &  1.440  &  144  \\
4   &  10226.0 $-$ 10252.3  &  0 $-$ 5  & 10 &  2.4  &  4.43  & 1.13  &  1.006  &  314  \\
5   &  10184.0 $-$ 10201.6  &  6 $-$ 11 & 9  &  2.0  &  1.12  & 0.54  &  1.009  &  183  \\
9   &  10230.5 $-$ 10250.5  &  7 $-$ 11 & 9  &  2.3  &  3.96  & 1.77  &  1.075  &  180  \\

\enddata 
\tablecomments{Col(1): The clump number; Col(2) The velocity range for the acceleration measurement; 
Col(3) The epochs of the spectra used for fitting (see also Table 3); Col(4) The number of Gaussian components that fit the data;
Col(5) The average linewdith of fitted features; Col(6) The best fit acceleration; Col(7) The uncertainty of the
acceleration; Col(8) The reduced $\chi^{2}$ of the fit; Col(9) The number of degrees of freedom of the fit.}
      
\end{deluxetable} 
\newpage

\subsubsection{Clump 9}

The masers in Clump 9 cover nearly the same velocity range as Clump 3, but the drifting of the whole maser pattern becomes much less clear because of both severe blending and reduced SNR. The blending causes the line structure to become very smooth and there are only two lines that can be seen to drift clearly in the time-velocity plot (Figure 2). For this clump, we fit nine Gaussian components with an average linewidth of 2.3 km~s$^{-1}$ to the masers between 10230.5 to 10250.5 km~s$^{-1}$. The measured acceleration is 3.96$\pm$1.77 km~s$^{-1}$~yr$^{-1}$ with a reduced $\chi^{2}$ of 1.075 (180 degrees of freedom). This is consistent with the value for clump 3, and implies that masers in clumps 3 and 7 are basically the same but seen at different times.

\subsection{Extended Radial Distribution of the Systemic Masers}
In Table 5 and Figure 3, one can see that different systemic maser features have different accelerations. In addition, we see evidence for some maser lines in Clump 8 that shows acceleration higher than 7 km~s$^{-1}$~yr$^{-1}$. Assuming that the maser accelerations are primarily due to the gravity of the black hole at the center of the maser disk, the 
acceleration values imply that the systemic masers arise at least at four different radii. This situation is in contrast to the systemic maser distributions in NGC 4258 (Herrnstein et al. 1999; Humphreys et al. 2008) and UGC 3789 (Paper II), where most systemic maser features reside either at one (NGC4258) or two (UGC 3789) narrow rings. Since the systemic masers in NGC 6264 have more extended radial distribution and the number of maser spots at each radius is small, it is not possible to the distance with the simple ``ring method"\footnote{The ``ring method" for distance determination only requires a single value of acceleration for each ring of systemic masers, the curvature of the rotation curve of high velocity masers, and the slope of the straight line that fits the position-velocity diagram of the systemic masers at each ring.} applied for UGC 3789. Therefore, 3-dimensional modeling of the warped disk as used in Paper IV is necessary and required to best utilize all observed quantities to obtain a precise Hubble constant and distance for NGC 6264.

\section{Distance Determination for NGC 6264 and the Hubble Constant}
\subsection{The Bayesian Approach to Hubble Constant Determination} 

To determine the Hubble constant with the NGC 6264 data, we adopt a Bayesian approach to model the maser disk in three dimensions (Reid et al. 2012). The Bayesian technique can fit a maser disk in three dimensions with maser clouds distributed arbitrarily across the thin disk.

The fitting program uses the same code for modeling the maser disk in UGC 3789 (Paper IV). In this program, we fit the maser data including the East \& North offset ($\Theta_{x}$ \& $\Theta_{y}$) of the maser positions on the sky, and the observed radial velocities ($V_{op}$) and accelerations ($A$) of the maser spots (Table 2) with a three dimensional model of the maser disk. We construct the model by placing each maser spot at radius $r$ and disk azimuth angle $\phi$ (with $\phi$ $\approx$ 0$^{\circ}$ for systemic masers and $\phi$ $\approx$ 90$^{\circ}$ for redshifted masers) on an orbit around a point mass (presumably a black hole) $M$. The position of the point mass on the sky is ($x_{0}$, $y_{0}$). Our model can allow the masers to have eccentric orbits, with eccentricity \emph{e} and pericenter rotated in angle $\omega$ = $\omega_{0}$ + ($\partial \omega/\partial r)r$ with respect to our line of sight. 

The disk can be warped in two dimensions:  the inclination warp $i(r)$ and the position-angle warp $p(r)$, each of which is specified by 3 parameters:
\begin{equation} 
   i(r) ~=~ i_{0}~+~(\partial i/\partial r)~r~+~(\partial^{2} i/\partial r^{2})~r^{2}~,      
\end{equation}
\begin{equation} 
   p(r) ~=~ p_{0}~+~(\partial i/\partial r)~r~+~(\partial^{2} i/\partial r^{2})~r^{2}~,      
\end{equation}
where r is the radius of the orbit in milli-arcsec. One can choose how many of these warping parameters one should actually use in the fitting depending on the degree of warping. The recession velocity of the galaxy is $V_{0}$, and the velocity correction to the pure Hubble flow due to the galaxy's peculiar velocity (i.e. the velocity caused by local gravity) is $V_{p}$. Note that $V_{0}$ and $V_{p}$ are non-relativistic, optical definition, in the CMB frame; the model and the data velocities (shifted from the observed LSR frame to the CMB frame by subtracting 19 km~s$^{-1}$) were converted internally in the fitting program to relativistically correct values.

In choosing the set of parameters to model the maser disk, unlike what was done for modeling the maser disk in NGC 4258 in Herrnstein et al. (1999), we fit $H_{0}$ instead of using the distance to the maser disk ($D$) as one of the global parameters (see the appendix). The distance $D$ is calculated from $H_{0}$ and $V_{0}$ when needed for the model.



The program adopts a Markov chain Monte Carlo (McMC) approach (e.g. Geyer 1992; Gilks, Richardson \& Spiegelhalter 1996) to obtain the posteriori probability distribution function of the model parameters. In particular, this program uses the Metropolis-Hastings algorithm (e.g. Metropolis et al. 1953; Hastings 1970) to choose the Markov-chain trial parameters, and a global step-size factor in the fitting is used to scale parameter steps so that an optimal Metropolis-Hastings acceptance rate of $\approx$23\% (Gelman, Gilks, \& Roberts 1997) is achieved. The distribution of the trial parameter values provides directly the posteriori probability distribution function of the parameters. 

Optimum values of the model parameters were estimated from the \emph{posteriori} probability distributions marginalized over all other parameters. They were generated from a total of 10$^{7}$ McMC trials, obtained from 10 independent program runs starting with slightly different model parameters. For the actual fitting process, we refer the readers to Reid et al. (2012) for detailed information.

\subsection{Hubble Constant Derived from the Bayesian Fitting}

The top panel of Figure 5 shows that the degree of the inclination warp of the NGC 6264 maser disk is small and the position-angle warp is less prominent than seen in NGC 4258 (Herrnstein et al. 1999; Argon et al. 2007). Therefore, we fit the warp only to the linear term in Equations (1) \& (2). In the Bayesian modeling, we found that the fitting becomes unstable when the eccentricity parameters are turned on, and therefore we decided to fit circular orbits and re-assess the impact of orbital eccentricity when new data is available. Note that while the magnitude of eccentricity is still unclear for the NGC 6264 maser disk, it is most likely that the orbital eccentricity is small ($e<0.03$ for UGC 3789 (Paper IV); $e<0.01$ for NGC 4258 (Humphreys et al. in prep.)). 

In order to account for systematic uncertainties in our measurements of the position, velocity, and acceleration of the maser spots, we estimate ``error floors'' (i.e. systematic errors) of 8 $\mu$as and 16 $\mu$as for $x$ and $y$, respectively, 1.0 km~s$^{-1}$ and 0.3 km~s$^{-1}$ for velocities of systemic and high-velocity maser features, and 1.0 km~s$^{-1}$~yr$^{-1}$ for accelerations of high-velocity masers. For accelerations of systemic masers, the estimated error floors range from 0.3 to 0.7 km~s$^{-1}$~yr$^{-1}$, depending on each individual maser clump's degree of line-blending, which is the main cause of systematic uncertainty of the acceleration measurements for systemic masers. We add the error floors to the observational data in quadrature before fitting the data. 

All model parameters except for one (see Table 6) were given flat priors and hence could vary freely, constrained only by the difference between data and model. The only parameter with a constraining prior was the galaxy's peculiar velocity, $V_{p}$. Since we do not have the expected peculiar velocity from simulation of large scale cosmic flows as for UGC 3789 (Paper II \& IV), we adopt $V_{p}=$0$\pm$300 km~s$^{-1}$ for NGC 6264 as an estimate.

Figure 5 shows the qualities of the Bayesian fitting by comparing one of the best fit models with the observed maser distribution, position-velocity diagram, and acceleration measurements. In general, the model matches the observations well, and the differences between the data and model are consistent with real uncertainties. The total reduced $\chi^{2}$ of the fit is 1.2 for 93 degrees of freedom.

\begin{figure}[ht] 
\begin{center} 
\vspace*{0.0 cm} 
\hspace*{-0.5 cm} 
\includegraphics[angle=0, scale=0.7]{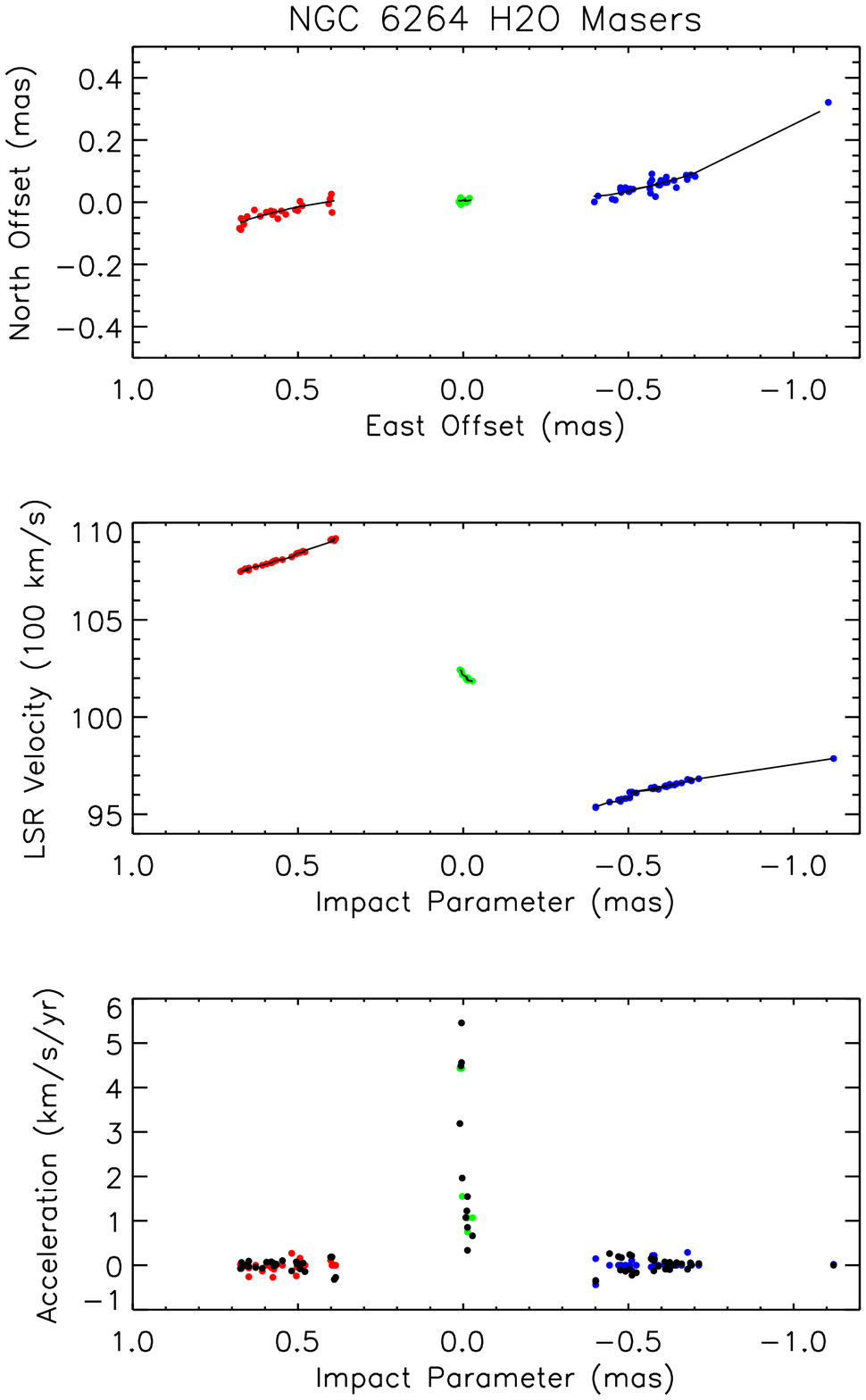}
\vspace*{0.0 cm} 
\caption{Data (colored dots) and best-fit model (lines and black dots). Top panel: Positions on the sky.
Middle panel: LSR velocity versus position along the disk. Bottom panel: Accelerations versus impact parameter. In all three plots,
the red, green, and blue dots show the redshifted masers, the systemic masers, and the blueshifted masers, respectively.}
\end{center} 
\end{figure}

The left panel in Figure 6 shows the maser distribution from overhead. As expected based on the range of measured accelerations, the systemic masers cover a large range of radii; the high velocity masers, with accelerations nearly zero, fall close to the mid-line of the maser disk. The latter feature supports the assumption made in Kuo et al. (2011) based on the NGC 4258 results (Herrnstein et. al. 1999) that the high velocity masers are close to the mid-line of the disk (within $\sim$13 degrees), and therefore the BH mass can be accurately determined by fitting a Keplerian rotation curve to the high velocity masers. In the right panel of Figure 6, we show the warp structure with the maser spots plotted on top of the warped disk model. Note that we decreased our viewing angle from 90$^{\circ}$ to 83$^{\circ}$ in order to show the warp structure more clearly.
   
\begin{figure}[ht] 
\begin{center} 
\vspace*{0.0 cm} 
\hspace*{-0.5 cm} 
\includegraphics[angle=0, scale=0.6]{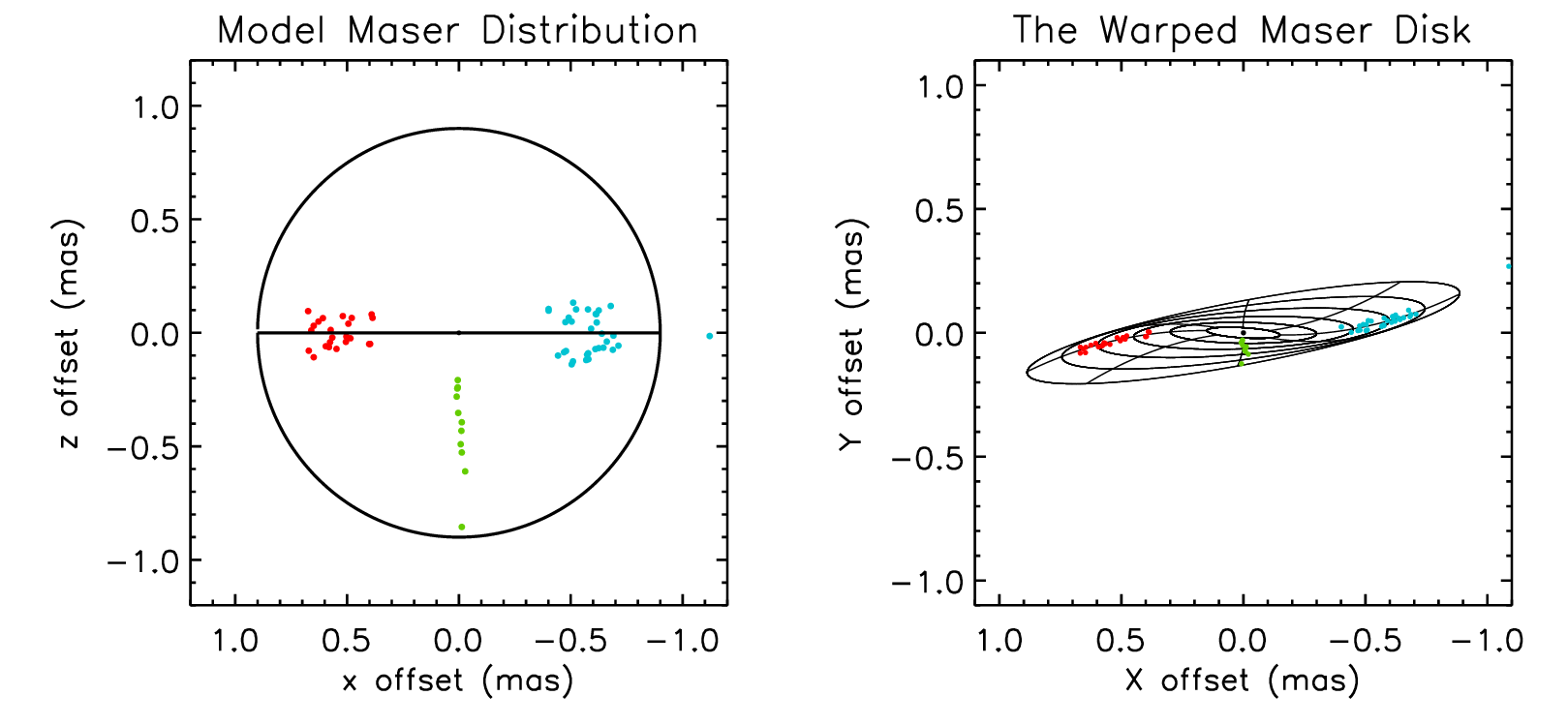}
\vspace*{0.0 cm} 
\caption{The left panel shows the model maser distribution in NGC 6264 from the overhead perspective. The right panel shows the best-fit model from the observer's perspective with model maser spots plotted on top of it. The red, green, and blue dots in both panels show the redshifted masers, the systemic masers, and the blueshifted masers, respectively. For illustration, we changed the observer's viewing angle from 90$^{\circ}$ to 83$^{\circ}$ to show the degree of disk warping more clearly.}
\end{center} 
\end{figure}    

We show the \emph{posteriori} probability distribution for $H_{0}$ from the Bayesian fitting in the left panel of Figure 7. The posteriori probability distribution is asymmetric about the peak of the distribution, which has a 68\% confidence range of $\pm$9 km~s$^{-1}$~Mpc$^{-1}$ centered at $H_{0} = ~$68 km~s$^{-1}$~Mpc$^{-1}$. Table 6 gives the best fitting values for all parameters. 

\begin{figure}[ht] 
\begin{center} 
\vspace*{0.0 cm} 
\hspace*{-0.5 cm} 
\includegraphics[angle=0, scale=0.6]{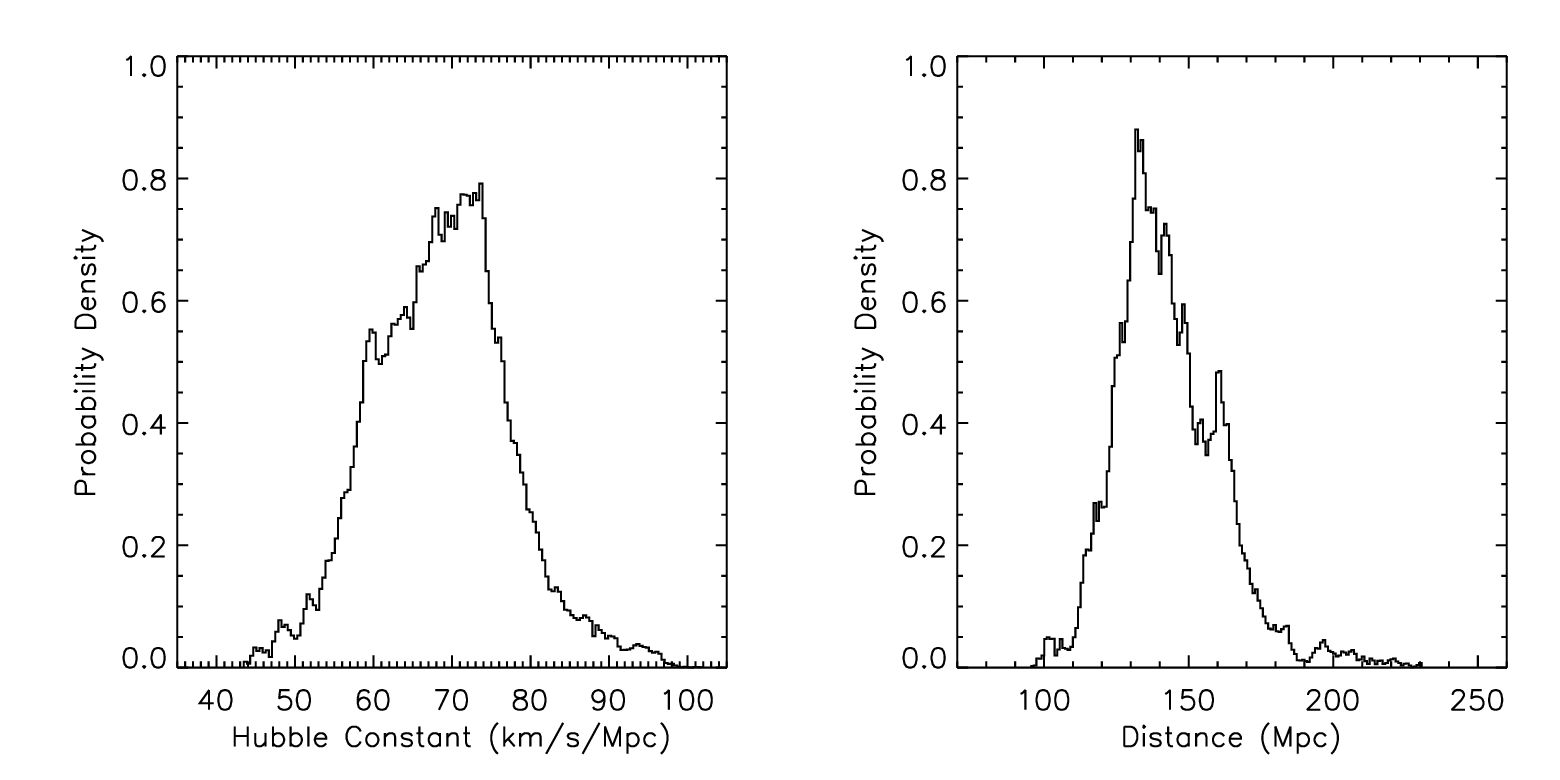}
\vspace*{0.0 cm} 
\caption{Marginalized posteriori probability distribution of the Hubble constant, $H_{0}$ (left panel), and the distance to NGC 6264, $D$ (right panel). The distribution for $H_{0}$ has a 68\% confidence range centered at $H_{0} = ~$68 km~s$^{-1}$~Mpc$^{-1}$ with an uncertainty of 9.0 km~s$^{-1}$~Mpc$^{-1}$. The distribution for $D$ has a 68\% confidence range centered at $D = $144 Mpc with an uncertainty of 19 Mpc.  }
\end{center} 
\end{figure}  

\begin{deluxetable}{cccccccccc} 
\tabletypesize{\scriptsize} 
\tablewidth{0 pt} 
\tablecaption{NGC 6264 H$_{2}$O Maser Model} 
\tablehead{ 
\colhead{Parameter}      & \colhead{Priors} & \colhead{Posterioris}   & \colhead{Units}  
}     
\startdata 
$H_{0}$ & ...  & 68$\pm$9       & km~s$^{-1}$~Mpc$^{-1}$    \\
$V_{0}$ & ... & 10189.6$\pm$1.2 & km~s$^{-1}$               \\
$V_{p}$ & 0$\pm$300     &  2$\pm$312        & km~s$^{-1}$               \\
$M$     &  ...          & 3.09$\pm$0.42      & 10$^{7}$~$M_{\odot}$      \\
x$_{0}$ &  ...         & 0.006$\pm$0.002    & mas                       \\
y$_{0}$ &  ...         & 0.007$\pm$0.003    & mas                       \\
$i_{0}$ &  ...          & 89.5$\pm$0.9         & deg                       \\
$\partial i/\partial r$ &  ...           & 0.7$\pm$2.4        & deg~mas$^{-1}$             \\
$p_{0}$ & ...      & 94.2$\pm$0.3       & deg                        \\
$\partial p/\partial r$ & ...      & 18.3$\pm$1.8       & deg~mas$^{-1}$              \\

\enddata 
\tablecomments{  
Parameters are as follows: Hubble constant ($H_{0}$ ), recession velocity
of central black hole in the CMB frame ($V_{0}$), peculiar velocity with respect to Hubble 
flow in cosmic microwave background frame ($V_{p}$), black hole mass ($M$), eastward
(x$_{0}$) and northward (y$_{0}$) position of black hole with respect to the average position
of the systemic maser features from the BB261K and BB261Q data, disk inclination ($i_{0}$) and
inclination warping (change of inclination with radius: $i_{1}\equiv di/dr$ ), disk position
angle ($p_{0}$) and position angle warping (change of position position
angle with radius: $p_{1}\equiv dp/dr$ ). Flat priors were used, except where
listed. Parameter values given in Table 6 were produced from binned histograms for each parameter and finding the center
and edge of the central 68\% of the probability distribution. We assign the difference between the center and edge
of the distribution, scaled by the square-root of the (reduced) chi-squared per degree of freedom, to be the parameter uncertainty. 
} 
\end{deluxetable}

\subsection{The Angular-diameter Distance to NGC 6264}

When modeling the maser disk, we use $H_{0}$ instead of $D$ as a fitting parameter (see the appendix). In order to obtain the posteriori probability distribution for the maser distance to NGC 6264, we first calculate the angular-diameter distance $D$ for each Markov-chain trial in the Bayesian modeling with the equation $D = (V_{0}+V_{p})/H_{0}$/(1+$z$), where $z$ is the redshift of NGC 6264. We then compute the marginalized posteriori probability distribution for $D$. The right panel of Figure 7 shows the posteriori probability distribution for $D$. The posteriori probability distribution of $D$ has a 68\% confidence range of 19 Mpc centered at $D = $144 Mpc.

\section{The Geometric Distance to A Galaxy Beyond 100 Mpc}

Traditionally, Hubble constant determinations have relied on indirect distance measurements 
through the extragalactic distance ladder, which is based on the period-luminosity relation 
of Cepheid variables. Direct distance determinations were limited to stars in the Milky Way, the Large Magellanic Cloud, and water masers in NGC 4258 (Distance $=$ 7.2$\pm$0.5 Mpc; Herrnstein et al. 1999). It is only recently that direct distance measurements to galaxies in the Hubble flow (Distance $>$ 30 Mpc) have been achieved (e.g. Braatz et al. 2010; Suyu et al. 2010; Reid et al. 2012).  

In this paper, we obtain a direct measurement of the \emph{angular-diameter} distance to a galaxy beyond 100 Mpc in a single step. NGC 6264 is  $\approx$20 times farther than NGC 4258, and the angular size of the maser disk is more than 10 times smaller. In addition, the maser flux densities are over 100 times fainter. The small size ($\sim$ 1.4 mas in diameter) of the maser disk makes it difficult to achieve high fractional accuracy of maser position measurements, and the low flux densities of masers have not only been a challenge for efficient VLBI imaging and accurate astrometry, but also make reliable acceleration measurements difficult. Moreover, instead of residing in a single narrow ring as in NGC 4258, the systemic masers in NGC 6264 are located at multiple radii. This prevents using the simple approach adopted for measuring the distance to UGC 3789 (Paper II); instead, we apply a more sophisticated 3-dimensional modeling for this megamaser disk to measure the Hubble constant and distance. 


\section{Summary} 
Our main conclusions are the following:

\begin{itemize}
\item[1.]
The application of the megamaser technique to galaxies deep in the Hubble flow is intrinsically more difficult because of significantly lower flux densities and smaller disk angular size. To image the maser disks in distant galaxies more efficiently, we developed a method to use multiple maser lines at different locations in the maser disk for performing self-calibration on weaker sources.

\item[2.]
The systemic masers in NGC 6264 are located at four or more radii from the black hole. 3-dimensional modeling that utilizes all information of maser spots including their positions, velocities, and accelerations is essential to measure the distance to NGC 6264 because of the complexity of its maser disks. 

\item[3.]
We modeled the H$_{2}$O maser disk in NGC 6264 with a Bayesian approach and obtained a Hubble constant of $H_{0} =$ 68 $\pm$9 km~s$^{-1}$~Mpc$^{-1}$, corresponding to an angular-diameter distance of 144$\pm$19 Mpc. The Hubble constant is consistent with the value obtained from UGC 3789 in Paper IV ($H_{0} = 69\pm7$). This is the first time the megamaser technique has been successfully applied to a galaxy beyond 100 Mpc.

\end{itemize} 
 
We expect that including new observations will enable a 10\% accurate estimate of the Hubble constant from NGC 6264, as well as constraining the magnitude of orbital eccentricity in the maser disk.

\appendix 
\section{Correlation Between Distance or H$_{0}$ and Black Hole Mass Estimates}

\begin{figure}[ht] 
\begin{center} 
\vspace*{0.0 cm} 
\hspace*{-0.5 cm} 
\includegraphics[angle=0, scale=0.55]{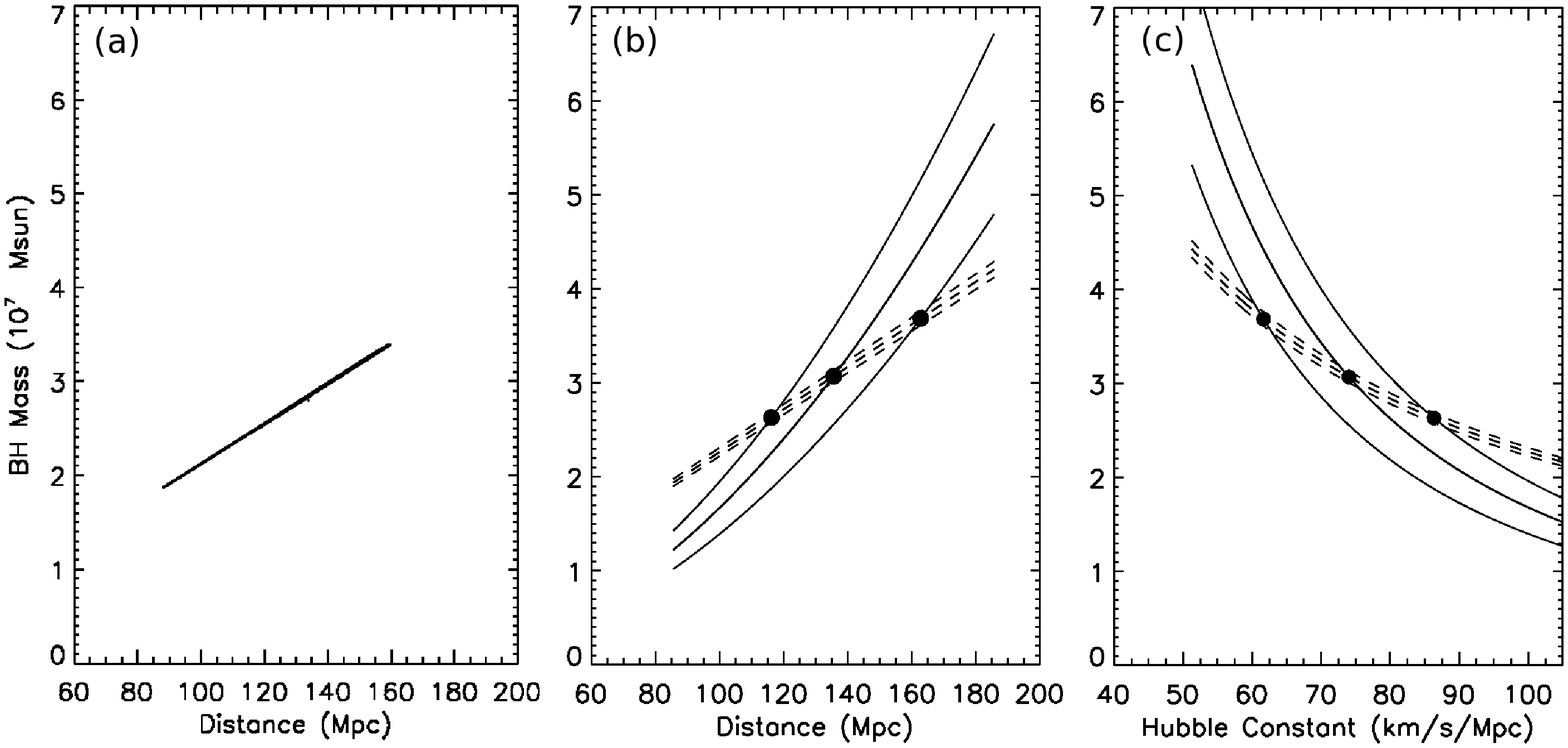}
\vspace*{0.0 cm} 
\caption{We illustrate the apparent correlation between $D$ and $M$ and the possible cause. The left panel shows the fitting results for galaxy distance $D$ and BH mass $M$ from our initial version of the Bayesian fitting program plotted on the $D-M$ diagram. The nearly straight-line appearance shows that $D$ and $M$ are highly correlated, and the correlation coefficient is nearly unity. The plot shown in the middle panel provides an explanation for this correlation. The black curves represent the quadratic relation between $D$ and $M$ from Equation (A2) and the dashed straight lines show the linear relation for $D$ and $M$ from Equation (A3) for a given set of measurements of $a_{sys}$, $k$, and $\Omega$. For a given distance, the black curves (from top to bottom) give the $M$ measured with $a_{sys}$, $k$, and $\Omega$, plus its 1 $\sigma$, 0 $\sigma$ and -1 $\sigma$ uncertainty. Likewise, the dashed lines (from top to bottom) give the measured $M$ based on Equation (A3) plus its 1 $\sigma$, 0 $\sigma$, and -1 $\sigma$ uncertainty. Every intersection of a dashed line and a black curve (the black spots) represents a solution for $D$ and $M$ from fitting the maser disk. As described in the appendix, we conclude that it is because the uncertainty in $k$ is usually significantly smaller than $a_{sys}$ and $\Omega$ that all solutions fall nearly on a straight line on the $D-M$ diagram. The presentation in the right panel is similar to that of the middle panel, with the only difference that $D$ in Equation (A2) and (A3) is replaced by $v_{sys}/H_{0}$, and we plot the Hubble constant on the x-axis. It can be seen that the solutions no longer fall on a straight line, and $H_{0}$ and $M$ are not perfectly correlated.}
\end{center} 
\end{figure}   

In our Bayesian modeling of maser disks in the MCP, we fit $H_{0}$ directly, rather than fitting $D$ as in Herrnstein et al. (1999). There are four reasons that make $H_{0}$ preferable over $D$ in the disk modeling: (1) since $H_{0}$ is the primary parameter we intend to measure, it is more convenient to save an extra step of calculation and have the posteriori probability distribution of $H_{0}$ directly output from the modeling program; (2) in this way the effect of uncertainty in peculiar velocity of a galaxy can be directly incorporated in the $H_{0}$ calculation; (3) since $D \propto$ 1/$a_{sys}$ and $a_{sys}$ usually has the largest fractional uncertainty among our observables, the posteriori probability distribution function of $D$ would usually be asymmetric; finally, (4) if we use D as a fitting parameter, its posteriori probability distribution is usually more highly correlated with $M$ than $H_{0}$ (See Figure 8a).

Superficially, a high correlation between D and M suggests that these parameters might be nearly degenerate.  This issue was not addressed in earlier work on NGC 4245 (Herrnstein et al 1999).   We found that replacing $D$ with $H_{0}$ in Bayesian modeling partially resolves this issue.
Nonetheless, the high correlation between D and M deserves further investigation in order to understand its nature and whether the results from the work on NGC 4258 and other sources that fit the maser disks with D and M are limited in accuracy.  We shall show that this behavior is a general characteristic of methods that rely on different powers of combinations of parameters and useful results can be obtained even when a formal correlation coefficient approaches unity. In the following analysis, we use a "simple" maser disk with systemic masers lying in a single narrow ring to demonstrate the case.

In the problem of modeling the maser disk and determining the angular-diameter distance, the relationship between $D$ and $M$ arises from two different measurements : the centripetal acceleration of the systemic masers and the Keplerian rotation curve for the high velocity masers. In the former case, we can see the relation between $D$ and $M$ in the equation for the centripetal acceleration $a_{sys}$:
\begin{equation} 
 a_{sys} = { GM \over r^{2} } = { GM \over D^{2}\Delta\theta^{2} },        
\end{equation}    
where $G$ is the gravitational constant, $r$ is the physical radius of the orbit of the systemic masers, and $\Delta\theta$ is the apparent angular radius of the systemic masers. In principle, $\Delta\theta$ can be obtained from the intersection points between the straight line traced by systemic masers (V $=$ $\Omega~\Delta\theta$) and the Keplerian rotation curve traced by the high velocity masers (V $=$ k~$\Delta\theta^{-1/2}$) in the position-velocity diagram of the maser disk (see Braatz et al. 2010). By solving for $\Delta\theta$ with the rotation curves for the systemic and high-velocity masers, we obtain $\Delta\theta = (k/\Omega)^{2/3}$. With this, we can re-write Equation (A1) and express $M$ as a quadratic function of $D$ :
\begin{equation} 
 M = { a_{sys} \over G } \biggl({k \over \Omega}\biggr)^{4/3}~D^{2}.        
\end{equation}  

On the other hand, since the coefficient $k$ of the Keplerian rotation curve is defined as k$\equiv$$\sqrt{GM/D}$, we can also find a linear relationship between $D$ and $M$ as
\begin{equation} 
 M = \biggl({k^{2} \over G}\biggr)~D.        
\end{equation} 

In Figure 8b, we plot Equation (A2) with black quadratic curves and Equation (A3) with dashed lines for a given set of measurements of $a_{sys}$, $k$, and $\Omega$ including their uncertainties. Every intersection of a dashed line and a black curve (black dots in Figure 8b) represents a solution for $D$ and $M$ from fitting the maser disk, and all possible solutions (within 1 $\sigma$ uncertainty) only fall in a region on the $D-M$ diagram whose shape is defined by the relative sizes of the uncertainties in $a_{sys}$, $k$, and $\Omega$. In current megamaser galaxies being observed for distance determination, the uncertainty in $k$ is usually small ($\sim$1\%) while the errors in $a_{sys}$ and $\Omega$ are more significant (e.g. $\ge$10\%). As a result, the dashed lines span a very narrow region on the $D-M$ diagram while the black curves cover a much broader region, and all the possible solutions for $D$ and $M$ fall on a region whose shape resembles a line. This makes $D$ and $M$ appear to be nearly 100\% correlated. Therefore, the apparent high correlation between $D$ and $M$ is not a result of intrinsic degeneracy, and thus the fitting results should be reliable.  

In Figure 8c, we demonstrate how the correlation between fitting parameters is reduced by replacing $D$ with $V_{sys}$/$H_{0}$ in Equations (A2) \& (A3). Now, one can see that the possible solutions no longer fall exactly on a straight line, and actually follow the $M$ $\propto$ 1/$H_{0}$ relationship. Because of this, the nearly perfect correlation doesn't exist between $M$ and $H_{0}$. Nonetheless, one can imagine that if  
the uncertainties in $a_{sys}$ and $\Omega$ are also small, then the black curves may cover a much smaller region, and the possible solutions may only fall in a small segment of the 1/$H_{0}$ curve. The distribution would thus resemble a straight line, and high parameter correlation appears again. This is indeed seen in our data, but because of the analysis shown here, high correlation should no longer be a serious concern.

\acknowledgements The National Radio Astronomy Observatory is a
facility of the National Science Foundation operated under cooperative
agreement by Associated Universities, Inc. We thank Ed Fomalont for
his kind help with our VLBI data reduction, and C.Y. Kuo thanks Mark
Whittle for his numerous insightful comments on our project. This research has made use of NASA's 
Astrophysics Data System Bibliographic
Services, and the NASA/IPAC Extragalactic Database (NED) which is
operated by the Jet Propulsion Laboratory, California Institute of
Technology, under contract with the National Aeronautics and Space
Administration.

\end{document}